\documentclass[12pt]{article}
\usepackage{graphicx}

\title{Functional-integral representation of atomic mixtures}

\author{O. Fialko and K. Ziegler \\
Institut f\"ur Physik, Universit\"at Augsburg, Germany}

\begin{document}

\maketitle

\begin{abstract}
A mixture of spin-1/2 fermionic atoms and molecules of paired fermionic atoms is studied in an
optical lattice. The molecules are formed by an attractive nearest-neighbor interaction.
A functional integral is constructed for this many-body system and analyzed in terms of 
a mean-field approximation and Gaussian fluctuations. This provides a phase diagram with 
the two merging Mott insulators and an intermediate superfluid. The Gaussian fluctuations give
rise to an induced repulsive dimer-dimer interaction mediated by the unpaired fermions.
The effect of an unbalanced distribution of spin-up and spin-down fermions is also discussed.
\end{abstract}



\section{Introduction}

A wide new field for investigating complex many-body systems has been opened
by the idea that clouds of atoms can be cooled to very low temperatures by
sophisticated cooling techniques \cite{phillips98}. 
Such an atomic cloud can be brought into a periodic potential which is created by
counterpropagating laser fields \cite{lewenstein}. This potential mimics the lattice
of core atoms of a solid-state system and is called optical lattice due to
its origin. It allows the simulation of conventional condensed-matter systems
as well as the creation of new many-body systems. New quantum states can
emerge due to the interplay of tunneling and interaction between the atoms.

In the following, a cloud of spin-1/2 fermionic atoms in an optical lattice
is considered, where an attractive interaction between atoms in nearest-neighbor
lattice wells is assumed. Our aim is to study different quantum phases that can
appear due to the formation and dissociation of molecules, and condensation
of the (bosonic) molecules.

\section{Model}

Our atomic cloud in an optical lattice is described by a grand-canonical system 
of spin-1/2 fermions
at temperature $1/\beta$ and chemical potentials $\mu_1$ and $\mu_2$, referring
to the two possible projections of the spin.
Its partition function is defined as a functional integral with respect to Grassmann
fields as \cite{ziegler, we}:
\begin{equation}
Z=\int e^{S(\psi,\bar{\psi})}D[\psi,\bar{\psi}] \ ,
\end{equation}
where $S$ is the action
\[
S(\psi,\bar{\psi})=\int_{0}^{\beta}
\left[\sum_{r}(\psi_{r}^{1}\partial_{\tau}\bar{\psi}_{r}^{1}+\psi_{r}^{2}
\partial_{\tau}\bar{\psi}_{r}^{2})-\sum_{r}(\mu_1\psi_{r}^{1}\bar{\psi}_{r}^{1}+
\mu_2\psi_{r}^{2}\bar{\psi}_{r}^{2}) \right.
\]
\begin{equation}
\left. -\frac{t}{2d}\sum_{\langle r,r^{\prime}\rangle}
(\psi_r^1\bar{\psi}_{r^{\prime}}^1+\psi_r^2\bar{\psi}_{r^{\prime}}^2) 
-\frac{J}{2d}\sum_{\langle
r,r^{\prime}\rangle}\psi_{r}^{1}\bar{\psi}_{r^{\prime}}^{1}
\psi_{r}^{2}\bar{\psi}_{r^{\prime}}^{2}\right]d\tau .
\label{model}
\end{equation}
$t$ is the tunneling rate of single fermions, whereas $J$ is the rate for
tunneling of a pair of fermions, located at nearest-neighbor sites in the optical
lattice. The $J$-term represents an attractive interaction.
In contrast to a local interaction, it provides a dynamics for the molecules. 
Depending on the ratio $t/J$, there is a competition between the individual
fermion dynamics (dominating for $t/J\gg1$) and the dynamics of molecules 
(dominating for $t/J\ll1$).

\section{Mean-field approximation and Gaussian fluctuations}

\begin{figure}
\begin{center}
\includegraphics[width=14cm]{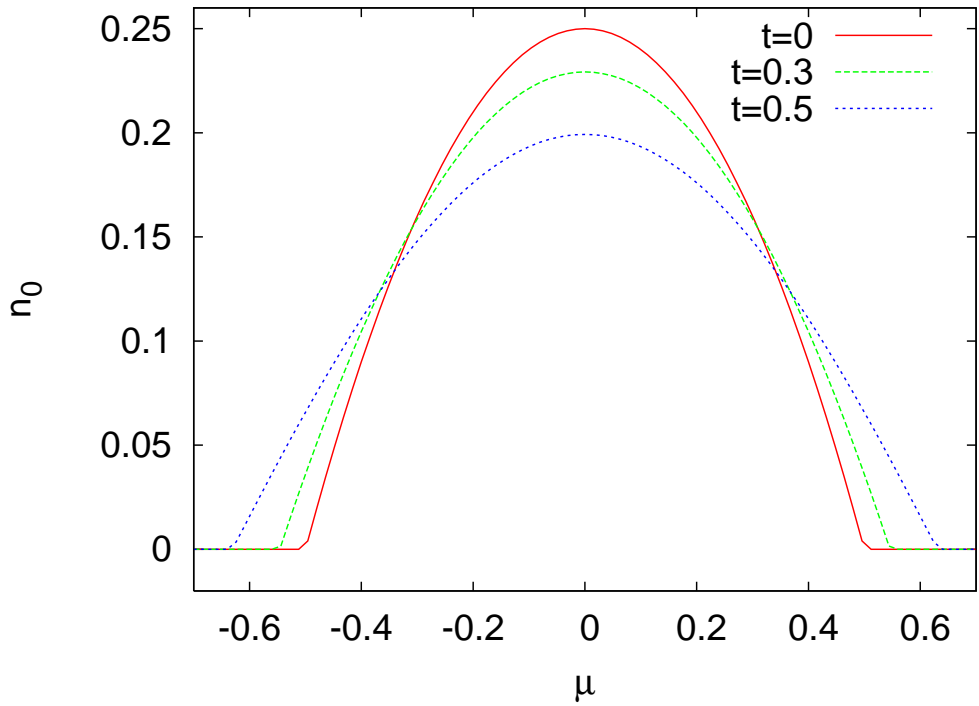}
\end{center}
\caption{Density of a molecular condensate for different single-fermion
tunneling rates $t$.
$\mu$ and $t$ are in units of $J$.}
\label{aba:fig1}
\end{figure}

\begin{figure}
\begin{center}
\includegraphics[width=14cm]{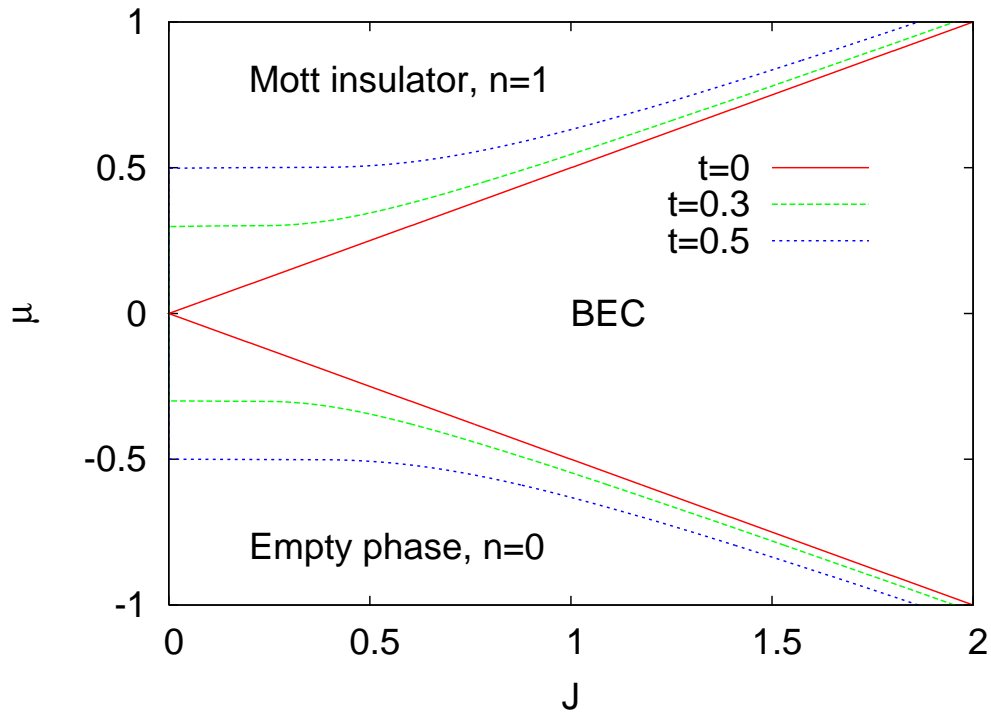}
\end{center}
\caption{Phase diagram for different values of $t$.
}
\label{aba:fig2}
\end{figure}

We decouple the $J$-term in the functional integral by two complex fields $\phi$ and $\chi$ 
\cite{ziegler,we}. $\phi$ is related to the order parameter for the formation
of molecules, and $\chi$ is required for stabilizing the complex integral. 
A subsequent integration over the Grassmann fields leads to
\begin{equation}
S_{\rm eff}=\int_{0}^{\beta} \left\{ \sum_{r,r^{\prime}}\bar{\phi}_r
\hat{v}_{r,r^{\prime}}^{-1}\phi_{r^{\prime}}+\frac{1}{2J}\sum_{r}\bar{\chi}_r\chi_r-\ln\det
\hat{\bf{G}}^{-1} \right\}d\tau \ ,
\end{equation}
where
\[
\hat{\bf{G}}^{-1}=\left(\begin{array}{cc} -i\phi-\chi &\partial_{\tau}+\mu_1+t\hat{w}  \\
\partial_{\tau}-\mu_2-t\hat{w} & i\bar{\phi}+\bar{\chi}  \end{array} \right),
\]
and a nearest-neighbor matrix $\hat{w}$ whose elements are $1/2d$ on the $d$-dimensional 
lattice. Moreover, we have $\hat{v}=J(\hat{w}+2\hat{1})$.  
The saddle-point (SP) approximation $\delta S_{\rm eff}=0$
for uniform fields and $\mu_1=\mu_2$
leads to the BCS-type mean-field result:
\begin{equation}
\chi=-\frac{2i\phi}{3},\ \ \frac{1}{J}=G,
\end{equation}
with
\begin{equation}
G=\frac{1}{\beta}\sum_{\omega_n}\int_{-1}^{1}\frac{\rho (x)}{|\phi|^{2}/9-(i\omega_n+\mu_1
+tx)(i\omega_n-\mu_2-tx)} dx \ ,
\end{equation}
and the density of states $\rho$ of free particles in the optical lattice. 
The mean-field calculations gives three phases: an empty phase, 
a Mott insulator and a Bose-Einstein condensate (BEC) of the molecules whose condensate 
density is
\begin{equation}
n_0=\frac{|\phi|^2}{9J^2}
\end{equation}
The latter is plotted in Fig. \ref{aba:fig1}, and the phase diagram is shown in 
Fig. \ref{aba:fig2}.

\begin{figure}
\begin{center}
\includegraphics[width=14cm]{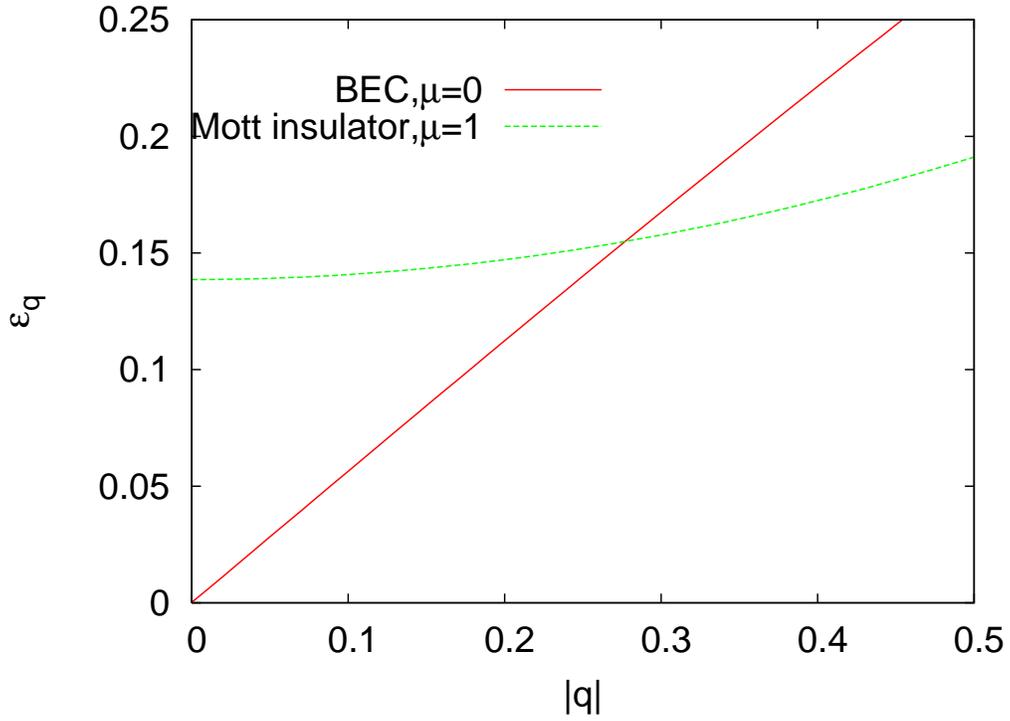}
\end{center}
\caption{Quasiparticle excitations for $t=0.5$ and $J=1$.}
\label{aba:fig3}
\end{figure}

Excitations out of the molecular BEC can be described by Gaussian fluctuations around
the SP solution by complex fields $\phi$ and $\chi$. Using $\Delta=i\phi+\chi$, 
$\bar{\Delta}=i\bar{\phi}+\bar{\chi}$, the corresponding action is
\begin{equation}
S_{\rm eff}=S_{\rm eff}^{0}+\delta S_{\rm eff}
\end{equation}
with
\begin{equation}
\delta S_{\rm eff}=\int_{0}^{\beta} \left\{ \sum_{r,r^{\prime}}\delta \bar{\phi}_r
\hat{v}_{r,r^{\prime}}^{-1}\delta \phi_{r^{\prime}}+\frac{1}{2J}\sum_{r}\delta \bar{\chi}_r
\delta\chi_r\right\}d\tau -\frac{1}{2}\mbox{tr}\left[\hat{\bf{G}}_{0}
\left(\begin{array}{cc}
-\delta\Delta & 0 \\ 0 & \delta\bar{\Delta} \end{array}
\right)\right]^{2}.
\end{equation}
The above result reads in terms of Fourier coordinates
\begin{equation}
\delta S_{\rm eff}=\sum_{q,\omega}\langle \delta\bar{\phi}_{q,\omega}, \hat{\bf{G}}^{-1}_{\rm 
eff}(q,i\omega)\delta\phi_{q,\omega} \rangle,
\end{equation}
where $\hat{\bf{G}}^{-1}_{\rm eff}$ is a 4 by 4 propagator, 
$\delta\phi$ is a four-component spinor. More details of the calculation can be found in Ref. \cite{we}. The excitation spectrum 
$\epsilon_q\equiv i\omega(q)$ is the solution of
\begin{equation}
\det\hat{\bf{G}}^{-1}_{\rm eff}(q,i\omega)=0.
\end{equation} 
This gives the Bogoliubov spectrum in the molecular BEC and a gapped 
spectrum outside the BEC (see Fig. \ref{aba:fig3}).

\section{Discussion and Conclusions}

\begin{figure}
\begin{center}
\includegraphics[width=14cm]{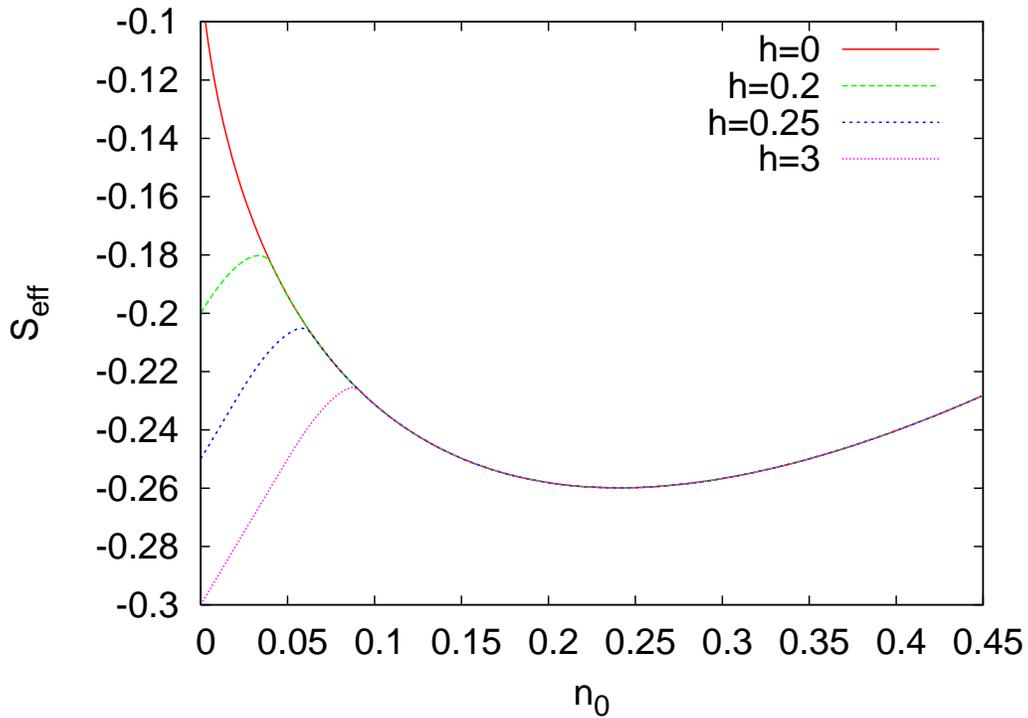}
\end{center}
\caption{Mean-field action of an unbalanced system, indicating a first-order
transition from the molecular condensate 
with increasing $h=(\mu_1-\mu_2)/2$. For $t=0.2, \mu=0$ and zero temperature 
the transition takes place around $h=0.25$.}
\label{aba:fig4}
\end{figure}

Our mean-field approach also allows us to consider an unbalanced molecular condensate
\cite{ketterle,hulet} 
by imposing different chemical potentials for the two spin projections of the
fermions, $\mu_1=\mu+h$, $\mu_2=\mu-h$.
Although we cannot address questions about nonlocal properties (like phase 
separation \cite{Sarma,Rupak,Stoof}) directly within our mean-field approach, the effect 
of two chemical potentials provides interesting effects even in a uniform system.
In particular, the existence of a first-order phase transition, usually leading
to phase separation, can be studied with the mean-field action of the unbalanced system
\begin{equation}
S_{\rm eff}\sim \frac{|\phi|^2}{9J}-\frac{1}{\beta}\int_{-1}^{1}
\rho(x)\ln\left[\cosh\left(\frac{E_{+}(x)\beta}{2} 
\right)\cosh\left(\frac{E_{-}(x)\beta}{2} \right) \right]dx ,
\end{equation}
where
\begin{equation}
E_{\pm}(x)=-h\pm\sqrt{|\phi|^2/9+(\mu+tx)^2} \ .
\label{en}
\end{equation}
This gives a first order-phase transition due to two separated minima in the mean-field action
for small $\mu$ or for larger $\mu$, depending on $h$ and $t$ (for fixed $J$) 
(cf. Fig. \ref{aba:fig4}).

\begin{figure}
\begin{center}
\includegraphics[width=14cm]{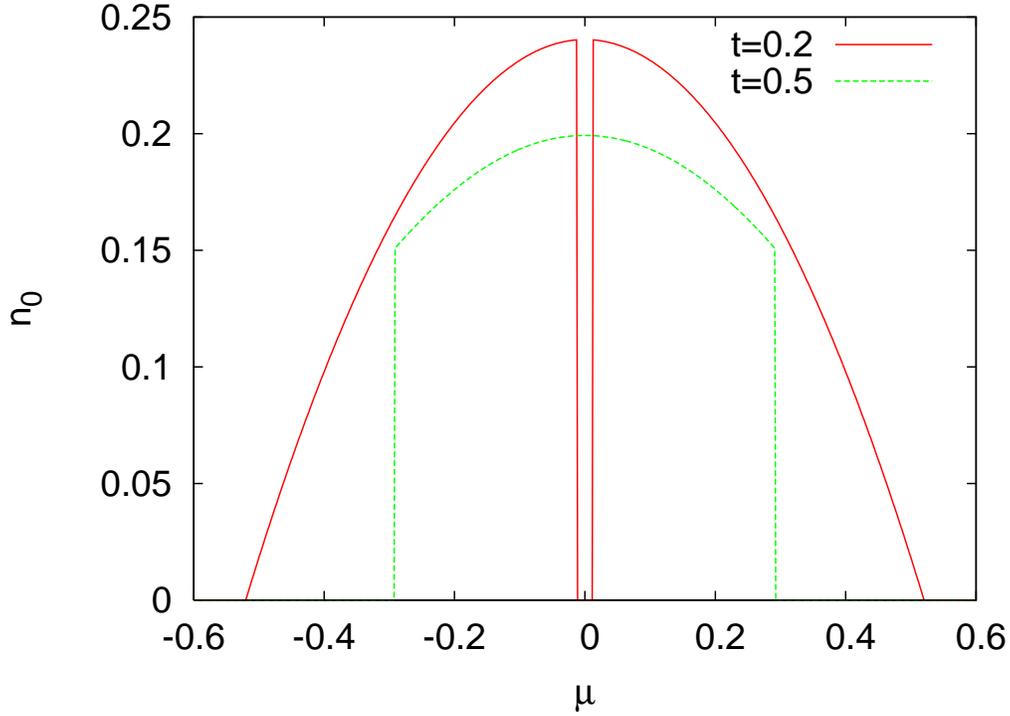}
\end{center}
\caption{Density of a molecular condensate as a function of 
$(\mu_1+\mu_2)/2$ for $h=0.26, J=1, T=0$.}
\label{aba:fig5}
\end{figure}

For small single-fermion tunneling rate $t$, there is a spin-polarized phase
simply because one spin projection has a negative chemical potential. This can
happen for small $\mu$ (cf. Fig. \ref{aba:fig5}). If $t$ is larger, there
is a shift of the chemical potential in Eq. (\ref{en}) by the single-fermion
tunneling rate. This prevents the appearence of a spin-polarized state for small
$\mu$ but it leads to a sudden disappearence of the molecular condensate
at large $|\mu|$, as shown in Fig. \ref{aba:fig5}.
This is accompanied by a first-order phase transition. 
At the point $h=\mu$  there might be a coexistence of molecules and spin polarized fermions 
\cite{Strinati,Griffin}.

In conclusion, our mean-field approach to the model of Eq. (\ref{model}) 
reduces to a BSC-type theory of molecules for spin-1/2 fermions with
attractive interaction.
It also provides a Mott insulator and a spin-polarized phase.   
The Gaussian fluctuations describe Bogoliubov-type excitations of a molecular condensate
and the gapped excitation spectrum of a Mott insulator. 

\end{document}